\documentclass[aps,prl,twocolumn,showpacs,superscriptaddress,
amsmath,amssymb,floatfix]{revtex4}
\usepackage{graphicx}
\usepackage{dcolumn}
\usepackage{bm}
\begin{document}
\title{Non-locality and short-range wetting phenomena}
\author{A. O. Parry}
\affiliation{Department of Mathematics, Imperial College 180 Queen's Gate,
London SW7 2BZ, United Kingdom}
\author{J. M. Romero-Enrique}
\affiliation{Department of Mathematics, Imperial College 180 Queen's Gate,
London SW7 2BZ, United Kingdom}
\affiliation
{Departamento de F\'{\i}sica At\'omica, Molecular y
Nuclear, Area de F\'{\i}sica Te\'orica, Universidad de Sevilla,
Apartado de Correos 1065, 41080 Sevilla, Spain}
\author{A. Lazarides}
\affiliation{Department of Mathematics, Imperial College 180 Queen's Gate,
London SW7 2BZ, United Kingdom}
\begin{abstract}
We propose a non-local interfacial model for 3D short-range 
wetting at planar and non-planar walls. The model is characterized by a 
binding potential \emph{functional} depending only on the bulk 
Ornstein-Zernike correlation function, which arises from different classes of 
tube-like fluctuations that connect the interface and the substrate.  
The theory provides a physical explanation for the origin of the effective
position-dependent stiffness and binding potential in approximate local
theories, and also obeys the necessary classical wedge covariance relationship
between wetting and wedge filling. Renormalization group and computer 
simulation studies reveal the strong non-perturbative influence of
non-locality at critical wetting, throwing light on long-standing theoretical
problems regarding the order of the phase transition.
\end{abstract}
\maketitle
 
Density functional \cite{Evans} and interfacial Hamiltonian models 
\cite{Forgacs} are complementary approaches to the theory of confined fluids. 
Mean-field, non-local density functionals give an accurate description of 
structural properties but are unable to account correctly for
long-wavelength interfacial fluctuations. To understand these it is usually 
necessary to employ mesoscopic interfacial Hamiltonians based on a 
collective coordinate $l({\bf{x}})$, measuring the local interfacial thickness.
These models are essentially local in character containing a surface energy 
term proportional to the stiffness $\Sigma$ of the unbinding interface and 
a binding potential function $W(l)$. In more refined theories the stiffness 
also contains a position dependent term \cite{FJ1}, $\Sigma(l)$, which, it 
is has been argued, may drive the wetting transition first-order \cite{FJ2}. 
Despite progress over the last few years there are a number of outstanding 
problems particularly for wetting with short-ranged forces. In addition, 
recent studies of fluids in wedge-like geometries have uncovered hidden 
connections or {\emph{wedge covariance}} relations between observables at 
planar wetting and wedge filling transitions \cite{Parry}, which have yet 
to understood at a deeper level. In this paper we argue that analogous to 
developments in density functional methods the general theory of short-ranged 
three-dimensional wetting should be formulated in terms of a 
{\emph{non-local}} (NL) interfacial Hamiltonian. The model we propose directly 
allows for bulk-like correlations arising from tube-like fluctuations 
\cite{Abraham} between the unbinding interface and the wall which contribute 
towards a binding-potential {\emph{functional}} $W[l,\psi]$ where 
$\psi({\bf{x}})$ is the shape function for the wall. Whilst the possible 
importance of such tube-like modes has been muted by several authors 
\cite{private}, to our knowledge this is the first quantitative treatment 
of them. Distinct contributions to $W[l,\psi]$ reflect the number of 
interacting tubes and have a simple diagrammatic interpretation. Three 
implications of the NL model are discussed: A) In the limit of small 
interfacial fluctuations the NL model identically recovers the known 
form of the local Hamiltonian allowing one to trace the specific position 
dependence of $W(l),\Sigma(l)$ to the Ornstein-Zernike (OZ) bulk correlation 
function. B) For filling in wedges with arbitrary tilt angles the model
obeys the classical wedge covariance relations observed in numerical studies 
of more microscopic theories. Such relations cannot be accounted for by local 
theories \cite{Parry2}. C) Renormalization group (RG) analysis show the 
{\emph{non-perturbative}} influence of NL interactions. 
Despite precise connection with the model and RG theory of Fisher and Jin (FJ)
at perturbative level the RG flow of the full NL model shows no stiffness
instability and the wetting transition remains second-order. Simulations 
confirm these findings.

\begin{figure}
\includegraphics[width=8.6cm]{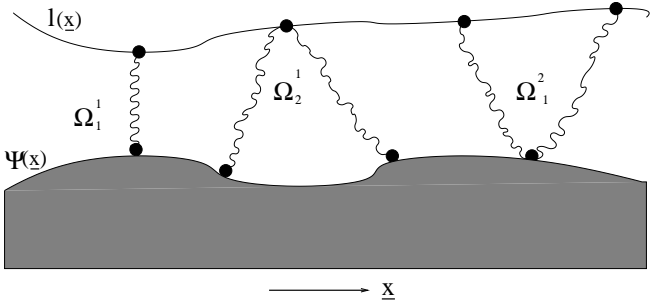}
\caption{Schematic illustration of the diagrams which represent the leading 
order contributions to $W[l,\psi]$. \label{fig1}}
\end{figure}

Consider a Landau-Ginzburg-Wilson Hamiltonian based on a continuum 
order-parameter (magnetization) $m({\bf{r}})$ in a semi-infinite geometry 
with bounding surface described by a single-valued height 
function $\psi({\bf{x}})$ where ${\bf{x}}=(x,y)$ is the parallel displacement 
vector. Denoting the surface magnetization by $m_1({\bf{x}})$ we write  
\begin{equation}
H_{LGW}=\int d{\bf{r}}\left(\frac{(\nabla m)^2}{2}+\phi(m)\right)+
\int ds_{\psi}\phi_1(m_1)
\label{LGW}
\end{equation}
where $ds_{\psi}=\sqrt{1+(\nabla \psi)^2}d{\bf{x}}$ is the wall
area element whilst $\phi(m)$ and $\phi_1(m_1)$ are suitable bulk
and surface potentials \cite{Nakanishi}. 
The bulk Hamiltonian is isotropic so the interfacial tension and stiffness 
are the same. 
Following FJ we identify the interfacial model $H=H_{LGW}[m_{\Xi}({\bf{r}})]$ 
where $m_{\Xi}({\bf{r}})$ is the profile which minimises Eq. (\ref{LGW})
subject to a given interfacial configuration. FJ determined 
$m_{\Xi}({\bf{r}})$ perturbatively in terms of local planar 
constrained profiles \cite{FJ1}. Here we construct 
$m_{\Xi}$ non-perturbatively using Greens' functions 
or equivalently correlation functions defined within the constrained 
wetting layer. The latter reduces to the classical OZ form over relevant 
distances \emph{provided} the wetting layer is \emph{much thicker} than 
its bulk correlation length $\kappa^{-1}$. The NL Hamiltonian is
\begin{equation}
H=\int d{\bf{x}} \left\{ \Sigma\sqrt{1+\left(\nabla l \right)^2}+
h(l-\psi)\right\}+W[l,\psi]
\label{nonlocal1}
\end{equation}
where $h$ is proportional to the bulk field. There is no
{\emph{explicit}} position dependent tension but rather a binding potential 
functional with three leading contributions 
\begin{equation}
W[l,\psi]=-a \Omega^1_1[l,\psi]+b_1\Omega^2_1[l,\psi]+b_2\Omega^1_2[l,\psi]
\label{nonlocalbinding}
\end{equation}
where $a$, $b_1$ and $b_2$ are best regarded as phenomenological parameters 
to be identified later. Each $\Omega_{\mu}^{\nu}$ represents integrated 
two-point interactions between $\mu$ and $\nu$ points on the wall and 
interface mediated by the (rescaled) bulk OZ 
correlation function $K(r)=\kappa e^{-\kappa r}/2\pi r$. These can be viewed as
contributions to the free-energy of a constrained thin-film arising from 
tube-like fluctuations of the bulk phase which tunnel from the interface to
the wall \cite{Abraham}. The first term involves only one tube 
\begin{equation}
\Omega^1_1[l,\psi]=\int ds_{\psi}^1\int ds_l^2  K(r_{12})     
\label{omega1}
\end{equation}
where $ds_l^{\mu}=\sqrt{1+(\nabla l({\bf{x_{\mu}}})^2}d{\bf{x_{\mu}}}$ etc and 
$r_{12}=\sqrt{\vert {\bf{x_{12}}}\vert ^2+
(\psi({\bf{x}_1})-l({\bf{x}_2}))^2}$ is the distance between two points on the
interface and wall. The last two terms 
\begin{equation}
\Omega^2_1[l,\psi]=\int ds_{\psi}^1\left\{ \int ds_l^2 
 K(r_{12})\right\}^2  
\label{omega2}
\end{equation}
\begin{equation}
\Omega^1_2[l,\psi]=\int ds_{l}^2\left\{ \int ds_{\psi}^1 
 K(r_{12})\right\}^2
\label{omega3}
\end{equation}
involve two tubes 
and may be viewed as a self-interaction between points on the same interface or
wall induced by the presence of a second surface. Each contribution can be
represented diagrammatically as shown in Fig.\ref{fig1}. 
The upper and lower lines represent typical non-planar configurations of 
the interface and wall. The undulated line joining them represents the 
interaction function $K(r_{12})$ whilst the solid dots imply integration 
over the area of each surface.
For general wall and interfacial configurations all contributions to 
$W[l,\psi]$ are NL. Simplifications arise when one or both are
planar. If both the wall and interface are flat, $\psi({\bf{x}})=0$, 
$l({\bf{x}})=l$ the Hamiltonian per unit area $W(l)=W[l,0]/A$ reduces to
\begin{equation}
W(l)=-a e^{-\kappa l}+(b_1+b_2)e^{-2\kappa l}
\label{bindingpotential}
\end{equation}
which recovers the standard form of the binding potential appearing in local
models. For the more general case of a non-planar interface near a planar 
wall, two contributions to the binding potential functional are local since
\begin{equation}
\Omega_\mu^1[l,0]=\int ds_l^1e^{-\mu \kappa l({\bf{x_1}})},{\indent}
\mu=1,2
\label{local1}
\end{equation}
However $\Omega_1^2$ remains NL and can be rewritten as a two-body 
repulsive interaction
\begin{equation}
\Omega_1^2[l,0]=\int \int ds_l^1 ds_l^2 S(\vert {\bf{x_{12}}}\vert;\overline l)
\label{Omegam1}
\end{equation}
where $\overline l=[l({\bf{x_1}})+l({\bf{x_2}})]/2$ and $S(x; \overline l)$ is:
\begin{equation}
S(x;\overline l)=\frac{\kappa^2}{2\pi}\int_{2 \kappa \overline l}^{\infty}
dt\frac{e^{-\sqrt{t^2+\kappa^2 x^2}}}{\sqrt{t^2+ \kappa^2 x^2}}\approx 
\frac{\kappa}{4\pi\overline l} e^{-2\kappa \overline l-\kappa x^2/4\overline l}
\label{w}
\end{equation} 
valid for $\kappa \overline l\gg 1$. In the {\emph{small gradient limit}}, 
$|\nabla l|\ll 1$, the NL term can be expanded and the model reduces to
\begin{equation}
H[l,0]\approx \int d{\bf{x}} \left\{ \frac{\Sigma(l)}{2} \left(\nabla l 
\right)^2+W(l) \right\}
\label{local5}
\end{equation}
with stiffness coefficient
\begin{equation}
\Sigma(l)=\Sigma-a e^{-\kappa l}-2b_1\kappa le^{-2\kappa l}+...
\label{local7}
\end{equation}
precisely recovering the FJ model and uniquely identifying $a$, $b_1$ and
$b_2$. In particular $a$ measures the deviation from the MF critical
wetting temperature, $b_2\propto a^2$ and the sign of $b_1$ determines the
order of the MF transition. Thus the origin of 
the $\kappa l e^{-2\kappa l}$ contribution, crucial 
in the FJ analysis, can be traced directly to a \emph{perturbative} 
treatment of the NL contribution $\Omega_1^2$. 

Now consider fluid adsorption in a wedge geometry ($\psi=\tan\alpha\vert 
x\vert$). The NL model satisfies the necessary requirement of 
{\emph{classical wedge covariance}} known from numerical studies of the 
microscopic model (\ref{LGW}) \cite{Parry2}. Classical wedge covariance 
refers to the relationship between observables at MF critical wetting 
and MF wedge filling transitions. Let $l_{\pi}(\theta)$ denote the 
MF thickness of the critical wetting layer written as a function 
of the contact angle. Let $l_w(\theta,\alpha)$ 
denote the  thickness of the filling layer above the wedge 
mid-point. Numerical minimization of the LGW Hamiltonian
(\ref{LGW}) shows that $l_w(\theta,\alpha)= l_{\pi}(\theta-\alpha)$ as 
$\theta\to \alpha$, for both shallow \emph{and} acute wedges. 
This relation cannot be explained using an approximate local Hamiltonian 
in which the interface interacts with the (closest) wall via a binding 
potential dependent on the normal distance \cite{Rejmer}. Such
models predict the \emph{incorrect} behavior $l_w(\theta,\alpha)= \sec\alpha
l_{\pi}(\theta-\alpha)$. In contrast the NL model obeys the correct 
wedge covariant relation. The reason for this can be traced to the structure 
of the NL binding potential. Since filling precedes wetting
($a\ne 0$), the dominant term is $\Omega_1^1$. Now for a 
flat interfacial configuration $l({\bf{x}})=l_0$ near a non-planar wall both 
$\Omega_1^1$ and $\Omega_1^2$ are local with, for example
\begin{equation}
\Omega[l,\psi]\vert_{l=l_0}=\int d{\bf{x}}\sqrt{1+(\nabla \psi)^2}e^{-\kappa
(l_0-\psi({\bf{x}}))}
\label{vert}
\end{equation}
showing that the effective local interaction occurs via the {\emph{vertical}}
distance to the surface. Near the filling phase boundary the interface is
essentially flat in the filled section of the wedge and the $\Omega_1^1$
contribution must be of the above form. This is sufficient to ensure
covariance. We also remark that for wetting at more general non-planar walls
the NL model reproduces the precise form of the stiffness matrix
appearing in approximate two-field models \cite{Boulter} 
valid for $\vert\nabla l\vert \ll 1$ and $\vert \nabla \psi \vert\ll 1$. 
This means that in application to complete wetting the NL theory satisfies 
exact sum rules \cite{Henderson}. 

Finally we turn to the controversy surrounding fluctuation effects at planar 
critical wetting. The standard capillary wave (CW) model, obtained by setting
$\Sigma(l)\equiv \Sigma$ in Eq. (\ref{local5}), famously predicts non-universal
criticality dependent on the wetting parameter $\omega=k_B T \kappa^2/4\pi
\Sigma$ \cite{FH}. However, this strongly disagrees with Ising model 
simulation studies \cite{Binder} which show only minor deviations from 
MF-like critical wetting behavior (for the experiments, see Ref. \cite{Bonn}). 
The more refined FJ model provides a 
possible explanation of this discrepancy since the $\Sigma(l)$ term drives 
the transition first order for physical values of $\omega$ \cite{FJ2}. 
Here we show that the stiffness instability is not a robust mechanism 
since the wetting transition described by the NL model remains 
continuous. A linear RG theory can be constructed provided we first expand 
$\sqrt{1+(\nabla l)^2}$ to square gradient order. The local terms 
$\Omega_\mu^1$ generate effective binding potential and position-dependent 
stiffness contributions which renormalize as in Refs. \cite{FJ2,FH}. 
We focus on the renormalization of the NL potential $S(x;\overline l)$ 
which controls the order of the phase transition since it is responsible for 
the $-le^{-2\kappa l}$ term in the perturbative $|\nabla l|\ll 
1$ limit. After renormalizing up to a scale $b=e^t$ the NL term
$\Omega_1^2$ retains its two-body form but with a modified potential 
$S_t(x;\overline l)$ satisfying the flow equation:
\begin{equation}
\frac{\partial S_t}{\partial t}=4 S_t + x\frac{\partial S_t}{\partial x}+
\omega \kappa^{-2} \left(\frac{1+J_0(\Lambda x)}{2}\right)
\frac{\partial^2 S_t}{\partial 
\overline l^2}
\label{difeq}
\end{equation}
where $J_0(x)$ is a Bessel function of first kind and $\Lambda$ is the 
momentum cutoff. This equation has the formal solution:
\begin{equation}
S_t(x;\overline l)=e^{4t} \int_{-\infty}^{\infty} dl \frac{\kappa S_0(xe^t,l)
\exp\left(\frac{-(\kappa l-\kappa \overline l)^2}{4\omega 
\Phi(\Lambda xe^t,\Lambda x)}\right)}
{\sqrt{4 \pi \omega \Phi(\Lambda xe^t,\Lambda x)}}
\label{formalsolution}
\end{equation}
where $\Phi(a,b)=\int^a_b dt [1+J_0(t)]/2t$. We choose $S_0(x,l)=\Theta (l) 
S(x,l)$, with $\Theta(l)$ the Heaviside step function and $S(x,l)$ given by 
Eq. (\ref{w}). 
As $t\to \infty$, $S_t(x,l)$ becomes increasingly localized around $x=0$. 
Using a matching technique we renormalize to a scale $e^{t^*}$ at which the 
curvature of the effective binding potential $W_t(l)$ at its global minimum 
is of order of $\Sigma \kappa^2$. Our $W_t(l)$ has a local 
contribution due to the $\Omega_1^1$ and $\Omega_2^1$ processes, and a 
NL contribution which is obtained from the expansion of $\Omega_1^2$ 
in powers of $\nabla l$: 
\begin{equation}
W^{NL}_t(l)=2\pi\int_0^\infty dx x S_t(x,l) 
\label{nlw}
\end{equation}
Numerical integration of the RG flow equation show that the wetting
transition is always \emph{second order}, and quantitatively similar 
to the non-universality exhibited by the CW model. This fact can be 
rationalized by noting that $\Phi(\Lambda x e^t,\Lambda x) \sim t$ 
as $t\to \infty$ and $\kappa x \lesssim e^{-t}$, which is the range where 
$S_0(xe^t,l)$ is non-negligible. Consequently, in our NL 
model there is \emph{no stiffness instability}. The difference with the RG 
predictions of the FJ model arises specifically from non-locality.
Mathematically the FJ flow equations can be recovered from Eq. (\ref{difeq})
if we approximate the Bessel function term by its quadratic expansion in $x$.
However this is not valid at large distances and invalidates the stiffness 
instability.

\begin{figure}
\includegraphics[width=8.6cm]{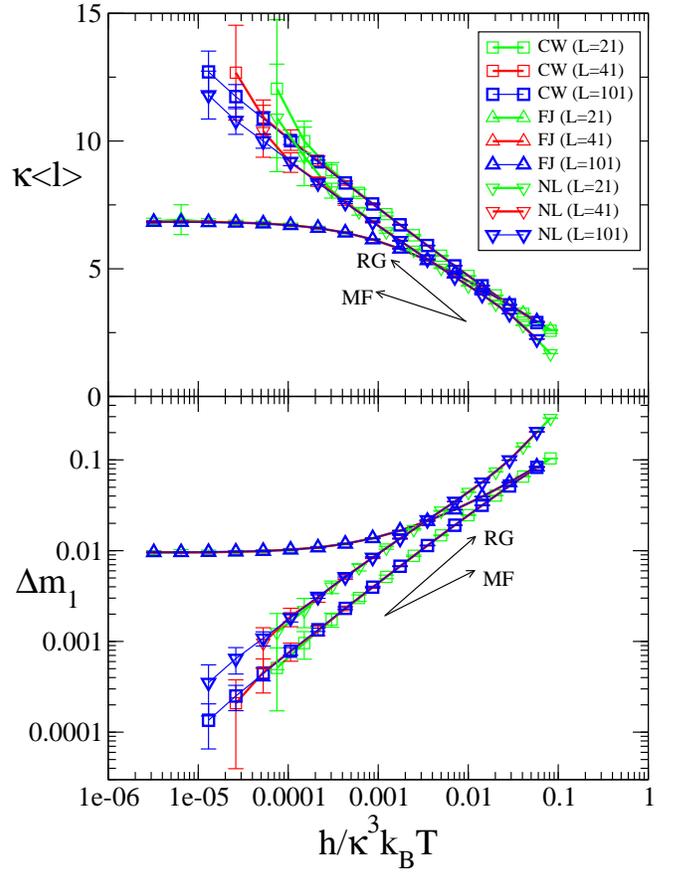}
\caption{(Color online). Plot of the mean wetting layer $\langle l \rangle$ and 
surface magnetization operator $\Delta m_1$ vs. $h$ obtained by computer 
simulations of the CW, FJ and our NL model for $\omega=0.8$, 
$a=b_2=0$ and $b_1/\kappa^2 k_B T=2.5$. 
\label{fig2}}
\end{figure}

In order to check the RG predictions, we have 
performed Monte Carlo simulations of the CW, FJ and NL Hamiltonians 
(with the approximation $\sqrt{1+(\nabla l)^2}\approx 1+ 
(\nabla l)^2/2$). Following Ref. \cite{Gompper} we discretize by introducing a
$L\times L$ lattice of spacing $\sigma$ with periodic 
boundary conditions in the directions parallel to the surface, but treating the 
interfacial position height as continuous variables. We chose
$\sigma=3.1623\kappa^{-1}$ so that $\Lambda \kappa^{-1}\sim \pi/\kappa 
\sigma \lesssim 1$, and also set $\omega=0.8$ and $b_1=2.5\kappa^2 k_B T$ 
which are reasonable Ising-like parameters. We anticipate the critical 
wetting phase boundary remains MF ($a=0$) for the CW and NL theories 
\cite{FH}, whilst the FJ exhibits a first-order transition at higher 
temperatures \cite{FJ2}. Fig. \ref{fig2} describes the behavior of the 
mean wetting layer thickness $\langle l \rangle$ and the surface 
magnetization-like operator $\Delta m_1=\langle e^{-\kappa l}
\rangle$ along the MF critical wetting isotherm $a=0$, $h\to 0$. 
The FJ model clearly describes partial wetting in this limit consistent
with a fluctuation-induced first-order transition. On the other hand the CW
and NL models are qualitatively similar, showing continuous wetting.
The divergence of the film thickness is well described by the RG result 
$\kappa \langle l\rangle
\sim -\sqrt{2\omega}\ln h$ even for moderately thick wetting layers.
However the surface magnetization shows a much larger preasymptotic
critical regime. The asymptotic non-universal behavior $\Delta m_1 
\sim h^{1-1/2\nu_\parallel}$, with $\nu_\parallel = 
(\sqrt{2}-\sqrt{\omega})^{-2}$ is \emph{not observed} until the wetting
layer $\kappa \langle l\rangle \sim 10$ for very large lattice sizes
$\kappa L\sim 300$. This is strongly suggesting that current Ising model
simulations will not be able to observe significant deviation from MF behavior
provided they focus on surface quantities. 

We conclude by mentioning extensions and limitations of our theory. It is 
straightforward to generalize the model to describe wetting at heterogeneous 
substrates with, for example, hydrophilic and hydrophobic domains \cite{Gau}. 
For such geometries \emph{all} the contributions to $W[l,\psi]$ will be 
NL and may influence mesoscopic wetting behavior. The same is true 
for other types of homogeneous sculpted substrates as, for example, wedges 
of parabolic cross section \cite{Parry3}. 
However, even for the simple case of wetting at 
homogeneous planar walls, systematic improvements of the theory can be 
envisaged. The binding potential functional (\ref{nonlocalbinding}) is 
only valid for wetting layers many times larger than $\kappa^{-1}$. If the 
wetting layer is only a few bulk correlation lengths thick, the interaction 
function $K(r)$ should be replaced by the correlation
function defined within the constrained profile. This would modify the
form of the binding potential functional at short distances and may 
influence how the asymptotic critical regime is approached.

A.O.P. gratefully acknowledges Prof. D. B. Abraham for discusssions regarding
the origin of the binding potential. We also thank Prof. 
M. E. Fisher and Dr. J. Gibbons for helpful comments. 
J.M.R.-E. acknowledges partial financial support from Secretar\'{\i}a de
Estado de Educaci\'on y Universidades (Spain), co-financed by the
European Social Fund, and from the European Commission under Contract
MEIF-CT-2003-501042. A.L. acknowledges partial financial support 
from the A. G. Leventis Foundation.

\end{document}